\documentclass[aps,preprint,showpacs,superscriptaddress]{revtex4}
\usepackage{graphicx}
\usepackage{amsmath,amssymb,latexsym}
\bibstyle{apsrev.bib}

\def\WIDTHE{\textwidth}  
\def\WIDTHD{0.9\textwidth}  
\newcommand{\beq}{\begin{equation}}
\newcommand{\eeq}{\end{equation}}

\begin{document}

\title{Knot localization in adsorbing polymer rings.}
\author{B. Marcone}
\affiliation{Dipartimento di Fisica, Universit\`a di Padova,
I-35131 Padova, Italy.}
\author{E. Orlandini}
\affiliation{Dipartimento di Fisica and Sezione CNR-INFM,
Universit\`a di Padova, I-35131 Padova, Italy.}
\affiliation{Sezione INFN, Universit\`a di Padova, I-35131 Padova,
Italy.}
\author{A. L. Stella}
\affiliation{Dipartimento di Fisica and Sezione CNR-INFM,
Universit\`a di Padova, I-35131 Padova, Italy.}
\affiliation{Sezione INFN, Universit\`a di Padova, I-35131 Padova,
Italy.}

\begin{abstract}
We study by Monte Carlo simulations a model of knotted polymer
ring adsorbing onto an impenetrable, attractive wall.
The polymer is described by a self-avoiding polygon (SAP) on
the cubic lattice.  We find that the adsorption transition temperature, 
the crossover exponent
$\phi$ and the metric exponent $\nu$, are the same as in the model
where the topology of the ring is unrestricted. 
By measuring the average length of the
knotted portion of the ring we are able to show that adsorbed
knots are localized.  This knot localization transition is triggered
by the adsorption transition but is accompanied by a less
sharp variation of the exponent related to the degree of localization.
Indeed, for a whole interval below the adsorption transition, one can
 not exclude a contiuous variation with temperature of this exponent. 
Deep into the adsorbed phase we are able to
verify that knot localization is strong and well described in terms 
of the flat knot model. 
\end{abstract}

\pacs{36.20.Ey, 64.60.Ak, 87.15.Aa, 02.10.Kn}

\maketitle

\section{Introduction}
Like other forms of topological entanglement of polymeric chains,
knots have relevant consequences for both physics and
 biology \cite{ImportantKnots}.
It is known that they can be found in long closed macromolecules
\cite{SoManyKnotsInTheBulk,ExpKnots1,ExpKnots2,OS07}, such as
circular DNA \cite{SoManyKnotsInTheBulkExp}, and that they can
affect important physical properties of them. For example, the
migration velocity of circular DNA in gel electrophoresis depends on
the knot type \cite{electrophoresis}. Among the various properties
of knotted polymers, the determination of the \emph{length} of the
knots inside them, i.e. the length of the part of the chain which
in some sense ``contains'' the entanglement responsible for the
overall knottedness, has attracted much attention in recent years
\cite{OurShortKnotPaper,OurLongKnotPaper,StasiakKnots,KardarTiraggio,VirnauMerdaCiCopiaDoloreEMorte}.
Indeed, the length of the knotted portion of the chain can be
expected to play an important role in determining its
 physical properties. For example, the diffusion coefficient of knots tied into DNA
by micromanipulation techniques should depend on the average size
of the knots \cite{QuakeVolo}. The action of topoisomerase on
knotted DNA surely depends on how localized the knot is
\cite{topoisomerasevsknots}.
The folding dynamics of a knotted protein \cite{knotsinproteins}
should also depend on the size of the knot.\\

The determination, and even the definition, of the knot length,
for real, 3D polymers, is however quite difficult. Thus,
historically, the problem was first faced for simplified models
in which one imagines that a knotted polymer ring is confined in
two dimensions. In this way the conformations of the ring reduce to
those of a polymer network in 2D. The network is made only of loops
and, under simplifying assumptions, maintains a fixed topology.
These objects are called \emph{flat knots}  \cite{FlatKnotsGuitterOrlandini,FlatKnots}.

One possible physical realization of flat knots, which allows
to trace a link between them and ``real'' knots, is the following:
think of a knotted polymer in 3D, and imagine it is fully adsorbed
on an attractive, flat surface; the polymer will then become two
dimensional, and will consist of loops, since the original 3D
chain was topologically a circle. The loops are made of segments
joined at vertices, which in general correspond to overlaps of the
adsorbed 3D polymer on itself. Usually, in the flat knot model the information
on the sign of crossings is not taken into account. 
One can study the length of the
``knot'' inside this object, at least under some simplifying
assumptions: namely, the number of crossings must be kept constant
and at the minimum value compatible with the corresponding topology in 3D. 
In fact, in such a case the network will consist of a fixed number $L$
of segments, while the knot length, $\ell$, can be unambiguously
put equal to the total length of the $L-1$ \emph{smaller} ones.
Now, one basic question about the knot length is whether knots are
\emph{localized} or not.
 Knots are said to be localized if, sending the polymer length $N$ to infinity, their (average)
length $ \langle \ell \rangle $ does not grow as fast as $N$. This
means that, in the \emph{thermodynamic limit} ($N \rightarrow
\infty$), the knot will
 behave as a point-like object with respect to the whole polymer.
More precisely, the localization can be of two types:
 $strong$ and $weak$. It is said to be $strong$ when
 $\langle \ell \rangle$ grows slower than any power of $N$ (e.g., as $\log(\log(N))$),
while it is $weak$ if  $\langle \ell \rangle \simeq N^{t}$, with
the exponent $t$ strictly less than 1 (but larger than 0). When a
knot is $delocalized$, $ \langle \ell \rangle $ grows
 as fast as $N$: in this case the knot
 will always occupy an extended part of the entire chain.
From the point of view of statistical mechanics, the determination
of the localization behaviour of knots is a most interesting issue,
since the  exponent $t$ is expected
 to be a \emph{universal} (model independent) quantity.
 Thus, it is not surprising
that research on knot length has focused on this aspect; this
is true for flat knots as well as for 3D knots, whose study
inherited some terminology, and some ideas, of flat knot theory.
For flat knots, Monte Carlo (MC) simulations, and theoretical
calculations which employ the theory of polymer networks
\cite{duplantier}, allow to make predictions on the value of $t$.
It turns out that  flat knots are \emph{strongly} localized in the
good solvent regime \cite{FlatKnots}, but undergo a delocalization
transition, and become delocalized,
below the $\theta$ point \cite{FlatKnots2,KardarFlatKnots2}. \\

The study of the localization behaviour of 3D knots is more
recent, and has been performed employing original strategies and
computer simulations for a consistent statistical definition of
knot length \cite{OurShortKnotPaper,OurLongKnotPaper,KardarTiraggio} . 
Indeed, the analytical treatment of the statistical
mechanics of polymers constrained to have the topology of
 a nontrivial knot is very
 hard \cite{theorydontwork}; this is mainly due to the \emph{non-local}
 character of the knottedness constraint, which makes
 impossible its description by a \emph{local} hamiltonian, and thus prevents, e.g., the use of
standard field-theoretical techniques. The Monte Carlo results
obtained in 
\cite{OurShortKnotPaper,OurLongKnotPaper}
 for $ \langle \ell \rangle $ as a function of $N$ have shown that prime
knots in 3D are \emph{weakly} localized, in the good solvent regime, with exponent $t \simeq 0.72$. The 
weak localization of 3D knots and the value of $t$ determined in \cite{OurShortKnotPaper,OurLongKnotPaper}
for SAPs on the cubic lattice have been subsequentially 
confirmed by simulations of off-lattice models \cite{VirnauMerdaCiCopiaDoloreEMorte}.

Since flat knots should describe fully adsorbed knotted polymers,
they are a useful model \emph{per se}, and not only for the
indirect, qualitative insights they provide into 3D knots.
Adsorbed polymers are in fact extensively studied
\cite{DeGennes,Vanderzande,JVRbook} and the adsorption transition
is an important paradigm of polymer statistics \cite{DeGennes}. \\
For polymer rings adsorbed on a plane it is known that knotting can
occur, and it has even been proven, for specific models, that it
occurs with probability 1 for infinitely long chains
\cite{AdsorbVanderzande,AdsorbJVR}. Thus, the behaviour of knots
should be expected to play an important role in determining the physical 
properties of adsorbed polymers, as it does for
swollen chains in 3D space. Flat knots represent only an extremely
schematic model of adsorbed knotted polymers. In fact, a realistic
model of an adsorbed polymer is given by a system consisting of a
3D polymer interacting with a short-range attractive,
impenetrable, plane. It is known \cite{AdsorbGrassb} that, in
the case of unrestricted topology, this
system exhibits a phase diagram with a desorbed phase at high
temperature $T$ and an adsorbed phase at low $T$; these two
regimes are separated by a phase transition at a certain critical
temperature $T_{c}$. At $T=0$ the polymer becomes a fully adsorbed
2D object, except possibly for $crossings$ (which are always
present if the topology of the polymer is different from that of
an unknot). Thus, the ground state of real adsorbed knotted
polymers should be described by flat knots; but for any nonzero
$T$, adsorbed polymers have $excursions$ (i.e. connected bunches
of desorbed monomers) which can be quite extended, even if they
have finite length on average.\\

So it is not clear if the true behaviour of real
adsorbed knots for $T>0$  should be similar to that of polymer networks of
fixed topology, or of their fully 3-dimensional counterparts, or,
maybe, of something in between. Recent measurements
\cite{Dietler}, on samples of knotted DNA adsorbed on a substrate,
indicate that knots are rather
localized. However, the chains in \cite{Dietler} seem not to
behave as fully 2D objects. For example, the measured $\nu$
exponent is between the 2D and the 3D value. Thus, a full
explanation of the behaviour of these polymers may require to go
beyond the simple flat knot model. The model of adsorbed polymers
we are going to study may be useful in this respect.

In this work we study the adsorption process of knotted ring
polymers by means of Monte Carlo (MC) simulations. We focus on the
adsorption transition and the low temperature (adsorbed) regime,
for the simplest prime knots ($3_{1}$, $4_{1}$, $5_{1}$ and
$5_{2}$). We check their thermodynamic properties, in order to
trace any significative difference between the behaviour of
 polymers with a fixed knot type and that of polymers with unrestricted topology.
We estimate the temperature dependence of the average knot length
$ \langle \ell \rangle $ and search for a possible transition
between the $T=\infty$ regime, where knots are expected to be
weakly localized, and the fully-adsorbed one ($T=0$), which
corresponds to flat knots and thus to an expected strong
localization. 

The paper is organized as follows. In Section II we describe the
model and the MC algorithms we use for the simulations. In section
III we present the numerical results and discuss the knot
localization properties in different regimes. We close this
section with a discussion on the relation between knotted polymers
in the strongly adsorbed regime and the model of flat knots.
Section IV contains our conclusions.


\section{Model and simulation methods}
A flexible polymer ring of $N$ monomers close to an impenetrable
surface can be modelled by an $N$-step self-avoiding polygon (SAP)
on the cubic lattice confined to the half-space $z \geq 0$ and
with at least one vertex anchored at the $z=0$ plane.  To include
a short-range attractive interaction  between the surface and the
polymer, an energy -1 is assigned to each vertex of the SAP having
$z=0$ (visit). Denoting by $v(\omega)$ the number of visits of a
given configuration $\omega$ the equilibrium properties of the
model are described by the partition function
\begin{equation}
Z_N(T) = \sum_{\{\omega \}}e^{v(\omega)/\kappa_B T}. 
\label{zeta}
\end{equation}
where  $T$ is the absolute temperature and $k_B$ is 
Boltzmann's constant. If the sum in Eq. (\ref{zeta}) extends to configurations
$\omega$ with all possile topologies, this model  displays, in the
thermodynamic limit, a second order phase transition from a
desorbed (high $T$) phase to an adsorbed one (low $T$)
\cite{DeGennes,Vanderzande}. In particular there exists $T_c > 0$
such that the  limiting free energy
\begin{equation}
{\cal F}(T) = \lim_{N\to\infty} N^{-1} \log Z_N(T)
\end{equation}
is equal to $-\log K^o_c$, independent of $T$, for all $T \ge T_c$
and is strictly greater than $-\log K^o_c$ for all $T < T_c$ \cite{HammerTorrieWhitting}. The limiting value $K^o_c$ denotes the critical fugacity of 
standard, non interacting 3D SAPs \cite{note1}. Let $\langle v
\rangle$ be the average number of visits and
\begin{equation}
\rho(T) = \lim_{N\to\infty}\frac{\langle v \rangle}{N}
\end{equation}
the limiting fraction of visits.  Then for
all $T > T_c$, $\rho(T)=0$ (desorbed phase) and for all $T< T_c$, 
$\rho(T) > 0$ (adsorbed phase).
 Right at the transition temperature $T=T_c\approx 3.497$ \cite{AdsorbGrassb}
 one expects
\begin{equation}
\label{crossover} \rho(T_c) \sim N^{\phi-1},
\end{equation}
where $\phi$ is the crossover exponent which is believed to be
very close to $1/2$ \cite{AdsorbGrassb}. Another way to detect the
adsorption transition is by looking at the metric exponent $\nu$
that controls the scaling of the radius of gyration through the
power law
\begin{equation}
\label{nu} \langle R_{g} \rangle \sim N^{\nu}.
\end{equation}
Indeed for $T\ge T_c$ one expects$\nu \approx 0.588$, i.e the value for 3D
SAPs \cite{Vanderzande} while for $T<T_{c}$ the value
$\nu = 3/4$ of  2D SAPs \cite{Vanderzande} should be recovered.\\

The adsorption transition is one of the most succesfully studied 
transitions in the polymer literature
\cite{Vanderzande,DeGennes}, but if one restricts the sum of the
partition function (\ref{zeta}) to SAPs with a given knot type
very few results are available so far. One of these is the
rigorous, strict, bound
\begin{equation}
{\cal F}^{\emptyset}(T) < {\cal F}(T) \qquad \forall T > 0
\label{zetao}
\end{equation}
where ${\cal F}^{\emptyset}(T)$ is the limiting free energy
(\ref{zeta}) restricted to the set of unknotted SAPs
\cite{AdsorbVanderzande,AdsorbJVR}. Inequality (\ref{zetao})
implies that for every finite value of $T$ the probability that
the polygon is knotted goes to one as $N$ goes to infinity
\cite{AdsorbVanderzande,AdsorbJVR}.  The only exception would be
the set of SAPs lying completely on the plane $z=0$ (2D SAPs) i.e.
the zero temperature limit for adsorbing rings. Of course, the
knotting probability for finite $N$ will depend on $T$ and this
has been investigated numerically \cite{AdsorbJVR}. No further
studies have been performed so far on the effect of topological
constraints on the adsorption transition but some reasonable
assumptions can still be made. For example, one should expect that
universal exponents such as the crossover exponent $\phi$ and the
metric exponent $\nu$ do not depend on topological constraints. 
On the other hand, the critical temperature $T_{c}$, could change 
when switching from the unrestricted
to a restricted topology ensemble, but there is no strong theoretical
insight on what should happen. Also, we cannot tell whether $T_{c}$ 
should depend on the specific knot type or not.
Given the difficulty in making
any progress in this problem by analytical or rigorous means a
natural way to gain insight is by Monte Carlo simulations.

For a fixed temperature $T$, SAPs with a fixed knot type are
generated by using a Monte Carlo approach based on the BFACF
algorithm, \cite{BFACF}. This is an algorithm which samples along
a Markov chain in the configuration space of polygons of variable
$N$ and with fixed knot type. The statistical ensemble considered
is thus grand canonical, with a fugacity $K$ assigned to each
polygon step. We adopt this algorithm because it preserves the
topology and is irreducible within each set of configurations
having the same knot type \cite{BFACFknots}. At a given $T$ we
used a multiple Markov chain (MMC) procedure \cite{Tesi} in which
configurations are exchanged among ensembles having different step
fugacities~\cite{Orlandini98}. This is done in order to improve
the efficiency of the sampling, especially at low $T$ where the
SAPs are strongly adsorbed on the plane. The temperatures we
considered are $T=3.50 \simeq T_{c}$ and 
$T=1.25$, $2.00$ and $2.75$, all much less then
$T_{c}$. We also consider the value
$1/T=0$ (non interacting case) in order to compare it with the
known situation of unweighted and geometrically unrestricted SAPs in 3D 
\cite{OurShortKnotPaper,OurLongKnotPaper}.

Despite the use of the MMC sampling technique, the BFACF algorithm becomes
quite inefficient in the strongly adsorbed phase and for high
values of $N$. To improve the sampling in this regime we decided
to use  an hybrid scheme based on a combination of the BFACF with
the pivot algorithm \cite{PivotBFACF}. Since any pivot move can
change the knot type of the resulting  SAP \cite{Madras}, a check
of its topology is needed before the move itself can be accepted.
This is done by calculating the Alexander polynomial $\Delta(z)$
in $z=-1$ and $z=-2$~\cite{Rolfsen,OldJETP}.

The BFACF algorithm does not preserve the value of $N$. 
Thus, in order to extract canonical
 averages at fixed $N$, we bin the data according to their $N$ value. 
To collect enough statistics for a given $N$ we used bins of
width 10. In this respect the symbol $\langle ..\rangle$ indicates for us
averages taken within a bin centered in $N$ and with size $10$.
The knot types considered in the simulations are the prime knots
$3_{1}$, $4_{1}$, $5_{1}$ and $5_{2}$. However, most of the
results we present here refer to the trefoil knot ($3_1$).

For each sampled SAP the length $\ell$ of the hosted knot $\tau$
is measured  by determining the shortest possible arc that
contains the knot.  The procedure works as follows
\cite{OurShortKnotPaper,OurLongKnotPaper}: given a knotted
configuration, open arcs of different length are extracted by
employing a recursive procedure. Each arc is then converted into a
loop by joining its ends at infinity (i.e., at very far dinstance) with a suitable path.  
The presence of the original knot is finally  checked by computing, on the
resulting loop, the Alexander polynomial $\Delta (z)$ in $z=-1$
and $z=-2$ (see \cite{OurLongKnotPaper} for
details). In all the simulations considered we sample, for each value of
$K$, over 20000 (independent) configurations. Since for fixed $T$
a MMC scheme with $10$ different $K$ values is used, the total
number of configurations considered in the statistics of a given
$T$ amounts to $2\times 10^5$.


\section{Results}

\subsection{Desorbed phase}
\begin{figure}[tbp]
\includegraphics[angle=0,width=\WIDTHD]{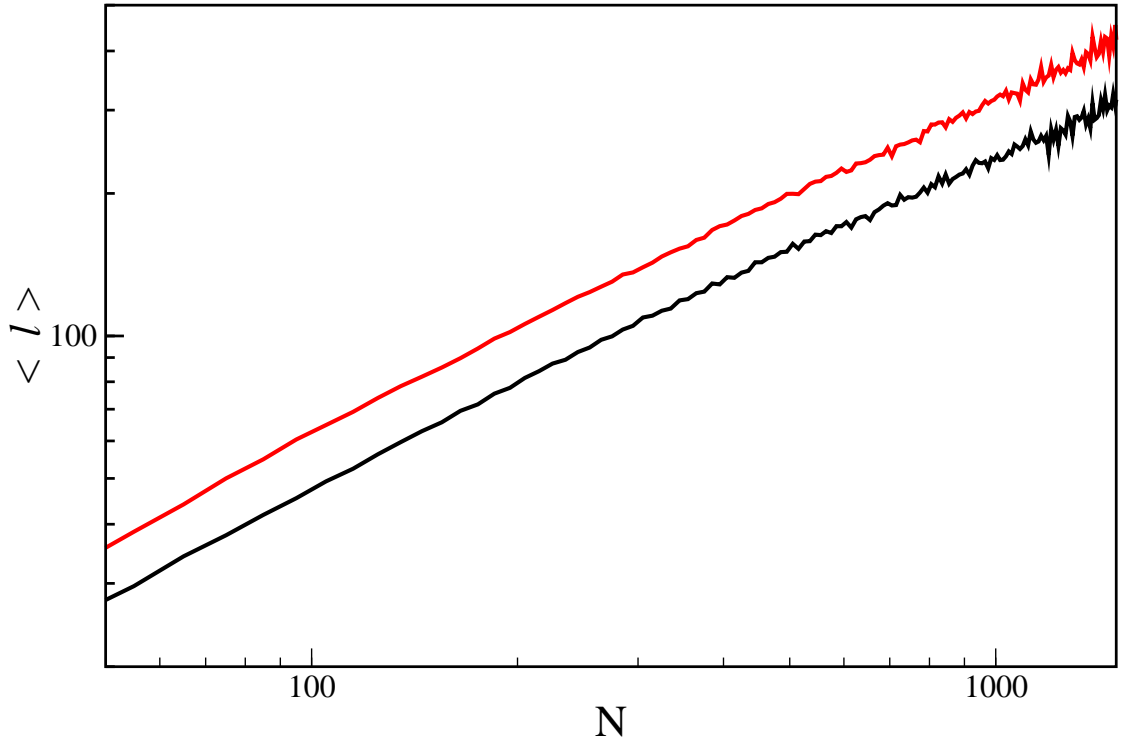}
\caption{(Color online) Average knot length $\langle \ell \rangle$
as a function of $N$ for SAPs in the bulk (bottom curve) and for
knotted SAPs confined in the $z\ge0$ half space (top curve). In
both cases the hosted knot is the trefoil knot ($3_1$). 
 $\langle \ell \rangle \simeq N^{t}$ holds in both cases with
$t=0.73\pm0.03$. This value is consistent with the estimated for 3D SAPs
 in the bulk \cite{OurShortKnotPaper,OurLongKnotPaper}.}
\label{fig1}
\end{figure}

To check the validity of our approach we first compare the known
situation of 3D swollen SAPs
\cite{OurShortKnotPaper,OurLongKnotPaper} with the one of SAPs
confined in the upper half space by an impenetrable non attractive
plane (non interacting case). Since the constraint $z \geq 0$
should not play a significant role in knot localization, we expect
strong similarities between the two cases. A first interesting
issue concerns the  the value of $K_c$ for the non interacting
and confined problem compared to the one ($K_c^o$) of the 3D case. For the
whole class of SAPs (unrestricted topology) it is known that
$K_c=K_c^o$ \cite{HammerTorrieWhitting} but no information is
available about the relation between $K_c(\tau)$ and $K_c^o(\tau)$
i.e. for the corresponding quantities for SAPs with fixed knot type $\tau$. 
By using a MMC with 10 different $K$'s ranging from $K=0.2109$ up to
$K=0.2130$~\cite{note2} we obtain a good evidence (within the
confidence limit and for the prime knots considered) that
$K_c(\tau)=K_c^o(\tau)=K_c^o$~\cite{notax}.
We also confirm that at $T=\infty$ the metric exponent $\nu$ coincides
(within error bars) in the two cases and that is
independent on the knot type (we estimate $\nu = 0.59 \pm 0.01$ for all knots
considered). 
We now turn out attention to the behavior of the
average knot length $\langle \ell \rangle$ as a function of $N$.
Previous studies have shown that for 3D swollen knotted SAPs
\begin{equation}
\label{scal} \langle \ell \rangle = A  N ^{t} + o(N^{t}),
\end{equation}
with  $t\approx 0.72$ for the trefoil
\cite{OurShortKnotPaper,OurLongKnotPaper}. Do we have the same
behavior if the $3_1$ SAPs are confined into one half space by an
impenetrable (but still not attractive) plane ? This seems to be
the case as witnessed by Fig. \ref{fig1} where a log-log plot of
$\langle \ell \rangle$ as a function of $N$ is reported for the
two situations. The two curves look indeed linear and parallel to
each other, confirming a weak localization regime with an exponent
$t$ that is the same (within error bars) in the two cases.

\subsection{Adsorbed phase}
When the attractive interaction between the SAP and the plane is
switched on, the entropy is not any more the only ingredient in
determining the equilibrium properties of the system. Instead, equilibrium
is determined by the interplay between entropy and the energy gain in flattening 
the polymer on the plane. Moreover, when the topology of the ring is restricted to
a fixed knot type, an additional entropic effect arises, because the entropy
of rings with fixed topology and of rings with unrestricted topology are different.

From a numerical point of view, simulations of SAPs in the
adsorbed regime require more effort than those for the desorbed
one. Let $K_c(T|\tau)$ be the critical fugacity for adsorbing SAPs 
at temperature $T$ and with knot type $\tau$.
For $T \ge T_c$ (desorbed regime), one can indeed assume, by
analogy with the unrestricted case, $K_c(T|\tau) = K_c(\tau)=
K_c^0(\tau)$. On the other hand, for $T<T_c$ (adsorbed phase),
$K_c(T|\tau)$ should decrease as the temperature decreases
\cite{note3} and one cannot rely any more on a known value of
$K_c$ for simulations at a given $T$. Hence, for each value $T<T_c$
considered the value of $K_c(T | \tau)$ must be estimated first
(by short MC runs) before collecting a significant amount of data for
that temperature.

\begin{figure}[tbp]
\includegraphics[width=\WIDTHE]{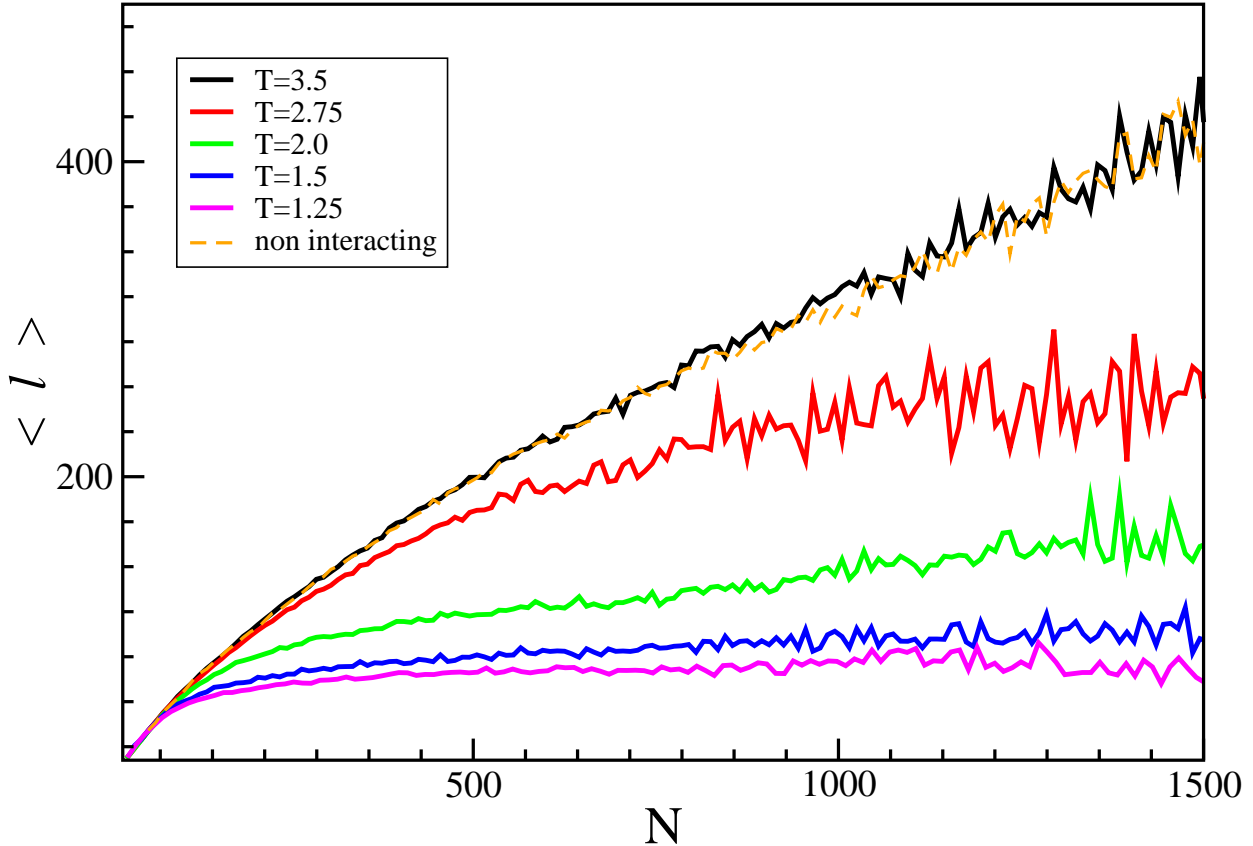}
\caption{(Color online) Average knot length  $\langle \ell
\rangle$ as a function of $N$ for SAPs with knot $3_{1}$. Different curves
correspond to decreasing (from top to bottom) values of the
temperature. The dashed curve corresponds to the non interacting
case. The corresponding values of $t$ can be deduced from the values of $c$
reported in Table II.} \label{lknotallT}
\end{figure}

Figure \ref{lknotallT} shows the average knot size $\langle \ell
\rangle$ as a function of  $N$ for SAPs with a  trefoil knot tied
in. Different curves correspond to different temperatures ranging
from $T=3.50$, a value just above the adsorption transition for the
unrestricted case, down to $T=1.25$, a value deep into the adsorbed
phase. The plot in the noninteracting case is also reported for comparison.
One can notice that the $N$ behavior of $\langle \ell \rangle$ for
SAPs close to the adsorption point $T=3.5$ coincides with the one
obtained for the non interacting case. Being the adsorption point
the last point of the desorbed phase, this result shows that in
this phase knots are weakly localized with exponent $t$ that does
not depend on $T$, and coincides with the one found for 3D
SAPs. Below the adsorption transition the situation changes
significantly: the average knot length still follows a power law
behavior (\ref{scal}) with $N$, but the exponent $t$ decreases as
$T$ decreases. More interestingly, if we go deeper into the adsorbed
phase ($T=1.5,1.25$) and look for  sufficiently large $N$,
$\langle \ell \rangle$ tends to a constant value. This
is a signal of a strong localization regime ($t=0$), 
reminiscent of the one found for flat knots~\cite{FlatKnots}.

Clearly if $t=0$, Eq. (\ref{scal}) does not give any insight into
the degree of localization of knots. More detailed information can
be however obtained by  analyzing the $N$ behavior of the
probability distribution function (PDF) of the knot length,
$P(\ell,N)$ \cite{OurLongKnotPaper}. In analogy with previous
works on similar problems \cite{Carlon,Zhandi,OurLongKnotPaper}
one can assume the following scaling form:
\begin{equation}
P(\ell,N) = \ell^{-c} \cdot g(\ell/N^{D}) \hbox{,} \label{scalP}
\end{equation}
where the scaling function $g$ is expected to approach rapidly
zero as soon as $\ell > N^D$, ($D \leq 1$). The quantity $N^D$ is
a cutoff on the maximum value $\ell$ can assume. We expect $D=1$,
because there is no reason \emph{a priori} to think that there
exists some `topological cutoff' which limits the size of the
knot. This is confirmed by our measurements, which yield $D \simeq
0.9 \div 1$ at every $T$. Assuming  $g$ is integrable when its
argument is sufficiently large, one can deduce that, for $0<t\leq
1$, $c=2-t$, while $c>2$ always implies $t=0$. In this respect the
desorbed phase (where $t\approx 0.75$) is characterized by
$c=1.25$ \cite{OurShortKnotPaper} , while at $T=0$ (fully adsorbed
polymer) we could expect  $c=2.69$ i.e. the value found for flat
knots \cite{FlatKnots}.

A common technique to analyze the scaling of the PDF (\ref{scalP})
goes as follows: for a trial value of $c$ and fixed $N$, one plots
$P(\ell,N)\ell^{c}$  versus $\ell/N^{D}$. Clearly, by varying $N$
different curves are displayed but if the values of $c$ and $D$ are close to
the correct ones, all these curves should collapse onto a single
one described by $g$. Eventually, after several trials,
``optimal'' values of $c$ and $D$ can be estimated. Unfortunately, to have
a good matching of the curves  an extremely good statistics is
required, and this would not be feasible in this context.

\begin{figure}[tbp]
\includegraphics[angle=0,width=\WIDTHE]{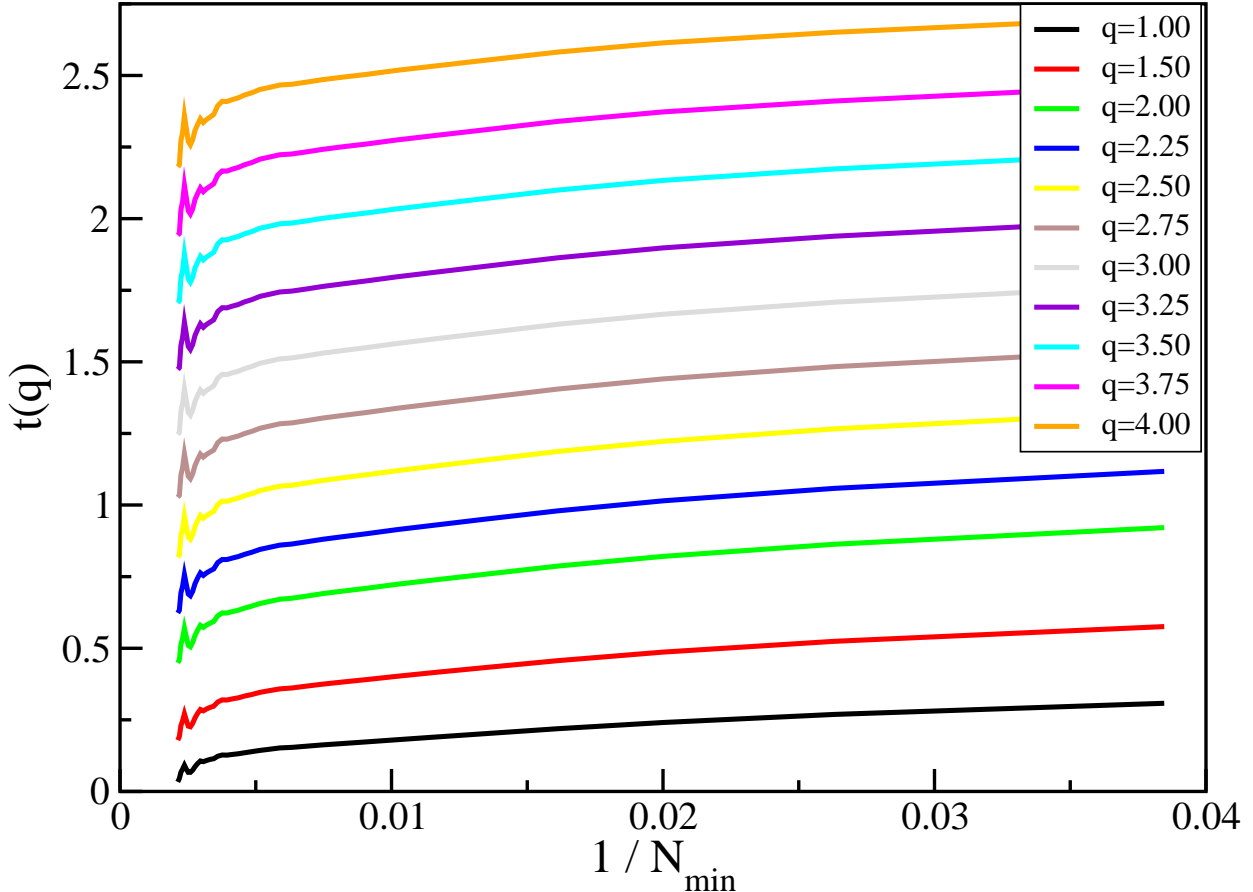}
\caption{(Color online) Finite-size scaling analysis of the
exponents $t(q)$ for $T=1.25$. For each value of $q$ ($q$
increases monotonically from bottom to top), $t(q)$ is obtained by
fitting the data of $\langle \ell^q \rangle$ as a function of $N$
with a power-law in the range between $N_{min}$ and
$N_{max}=1500$, with $N_{min} \ll N_{max}$. Here, different curves, corresponding to different
values of $q$ ranging from $q=1.00$ (bottom curve) to $q=4.00$
(top curve), are shown. To extract the asymptotic values we
compute $t(q)$ as a function of $N_{min}$ and extrapolate these
values as $1/N_{min} \rightarrow 0$: This gives our best estimate
of $t(q)$.} \label{fig:2}
\end{figure}

\begin{figure}[tbp]
\includegraphics[angle=0,width=\WIDTHE]{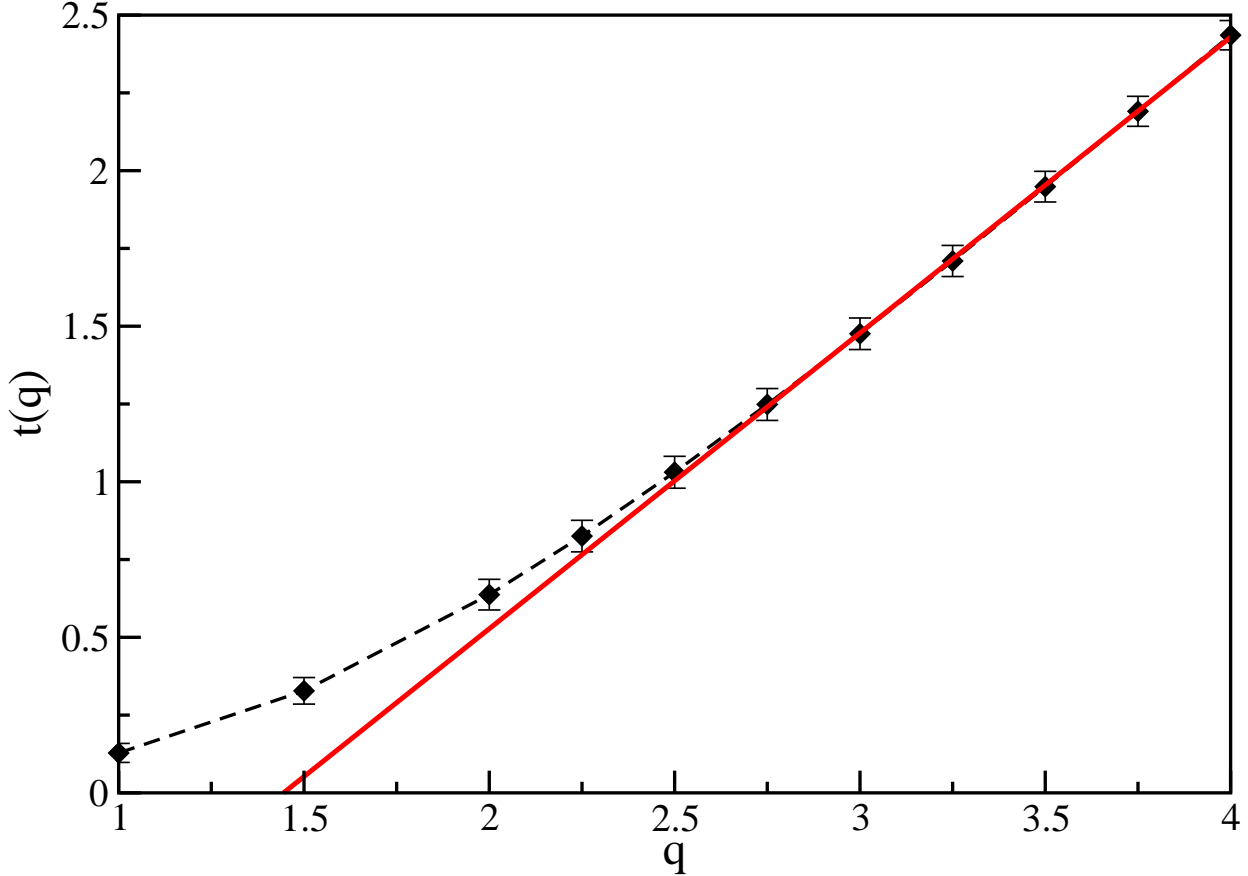}
\caption{(Color online) Analysis of the moments of the pdf
$P(\ell,N)$ for $T=1.25$ and for the knot $3_1$. The exponents
$t(q)$, calculated as described in the caption of the previous
picture and in the main text, are plotted against $q$. For $q \ge
2.5$  a good linear behavior is obtained and a linear fit in that
range of $q$ gives the estimate of $c$. } \label{fig:3}
\end{figure}

\begin{figure}[tbp]
\includegraphics[angle=0,width=\WIDTHE]{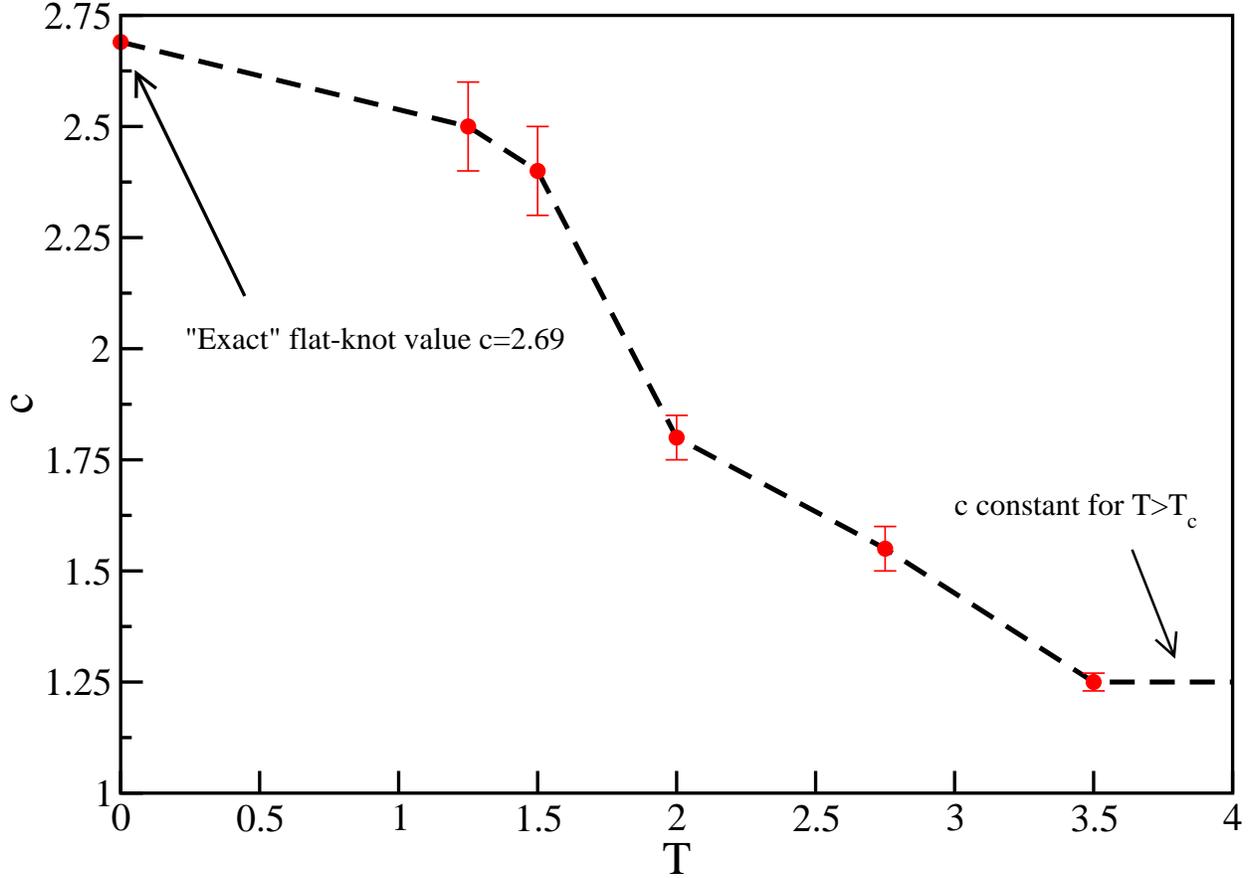}
\caption{(Color online) Estimated values of the  exponent $c$ for
SAPs at the adsorption transition ($T=3.5$) and in the adsorbed
phase ($T<3.5$). The knot considered is the trefoil. For $T \geq
T_{c}$, we get  $c=1.27 \pm 0.03$, independently on $T$ (desorbed phase) This
is consistent with $c=2-t$ and the expected value ($t\approx 0.75$)
found in this regime.} \label{tau_T}
\end{figure}

We can instead perform an analysis based on the scaling behavior
of the moments of the PDF in Eq. $(1)$ ~\cite{StellaMomentsTech}.
This method relies on the following consideration: given the
scaling behavior (\ref{scalP}) for the PDF, its $q$-th moment
$(q>0)$ should obey the asymptotic law:
\begin{equation}
\langle \ell ^ q \rangle = \int \ell^{q-c} g(\ell/N^{D}) \sim
N^{Dq+D(1-c)} \equiv N^{t(q)} \label{4}
\end{equation}
and the two parameters $D$ and $c$ can be deduced by fitting the
 estimated exponents $t(q)$ against the order $q$ \cite{note4}
and performing a finite size scaling analysis (see Figure \ref{fig:2}).
As shown in Figure \ref{fig:3} for the $T=1.25$ case, the plots of
$t(q)$ show deviations from linearity at relatively low $q$, due
to finite $N$ scaling correction effects. This is typical for this
kind of analysis \cite{StellaMomentsTech}. However, for a
sufficiently wide range of $q$,  a linear behavior can be
identified whose intercept gives an estimate  of $c$.
Repeating the above procedure for the different temperatures
considered one obtains the estimates plotted in Figure \ref{tau_T}
and reported in Table \ref{table1}.

For $T \geq T_{c}$ the value of $c$ is roughly $1.25$
\emph{independent on $T$}. This is consistent with the findings
presented in the previous section. Indeed, in the desorbed regime
(and up to the adsorption point included) the knot length exponent
$t$ is $\approx 0.72$. As the temperature is lowered the polymer
goes more deeply into the adsorbed phase and the $c$ exponent
increases reaching, at $T=1.25$, the value $c = 2.55 \pm 0.10$.
The curve in Figure \ref{tau_T} furnishes a good evidence that
knots in the adsorbed regime become strongly localized. The value of
$c$ at $T=1.25$ ($2.55$) agrees within the error bars with the
value $2.69$ found for flat knots, but it is possible that such
value is reached precisely only in the $T\to 0$ limit.

One could take the intersection between the curve in Figure~\ref{tau_T} and the
line $c=2$ as an estimate of a transition point $T_{loc}$ between 
the weak and the strong localization regimes. This would suggest a localization
transition occuring well below the adsorption transition.
On the other hand, the curve in Figure~\ref{tau_T} is an estimate 
of $c(T)$ that relies on finite $N$ simulations and it is hard 
to decide whether $c(T)$ would show a sharp discontinuity
as $N\to \infty$. 
Most intriguing would be the possibility of a range of temperatures 
in which the exponent $c$ varies with $T$. 

\begin{table}
\begin{center}
\begin{tabular}{|c|c|}
\hline
\bf{$T$} & \bf{c} \\
\hline
$3.50$  & $1.27 \pm 0.03$\\
$2.75$  & $1.55 \pm 0.05$\\
$2.00$  & $1.80 \pm 0.05$\\
$1.50$  & $2.42 \pm 0.10$\\
$1.25$  & $2.55 \pm 0.10$\\
\hline
\end{tabular}
\caption{Estimates of the exponent $c$ for different values of $T$
for trefoil knots. They have been obtained by the finite size
scaling analysis of the moments of the knot length as explained in
the text.} \label{table1}
\end{center}
\end{table}

\begin{figure}[tbp]
\includegraphics[angle=0,width=\WIDTHE]{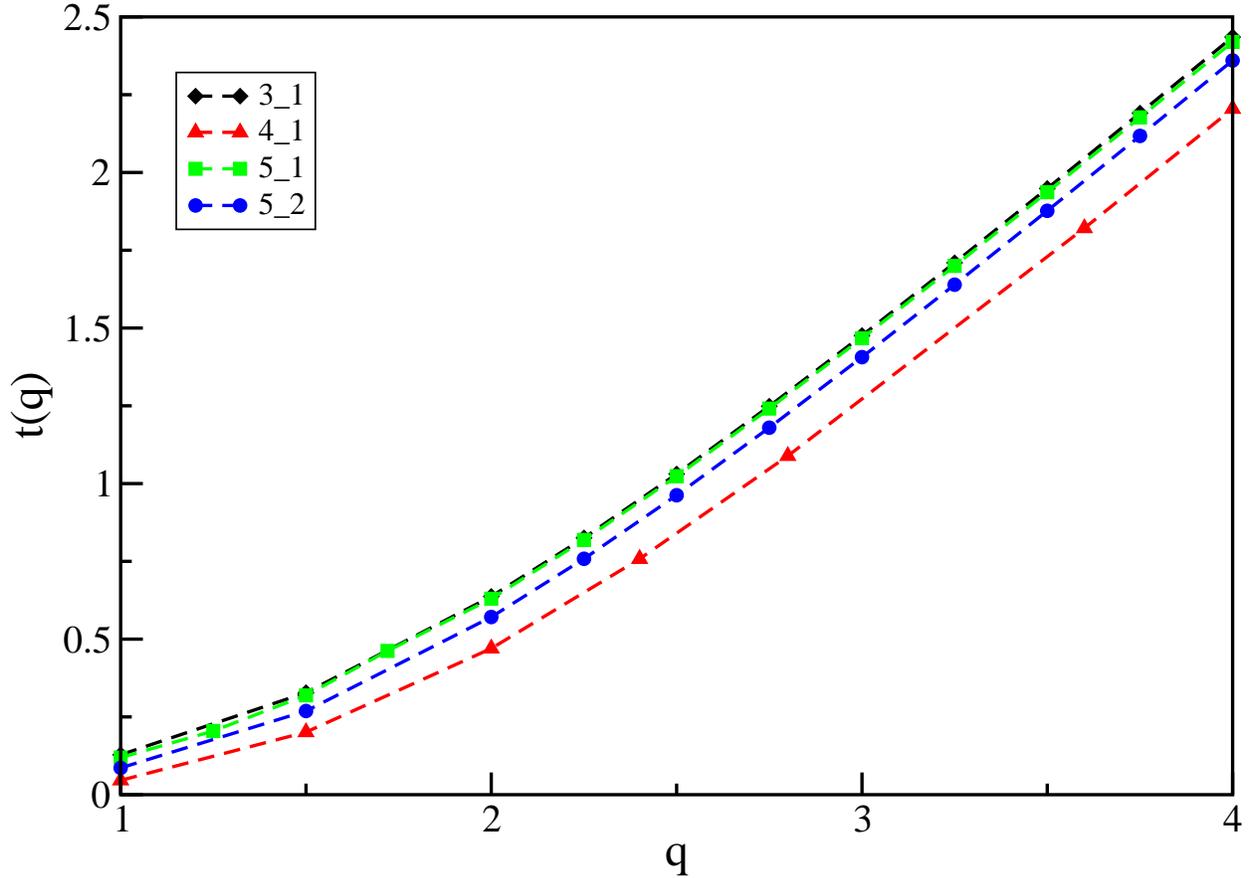}
\caption{Estimates of the exponent $t(q)$ as a function of $q$ for
$T=1.25$ and for different prime knots. The error bars, that are
not reported here for clarity, are of the same order as the ones
reported for $3_1$ in figure \ref{fig:3}.} \label{fig:7}
\end{figure}

Results for other prime knots are quite similar to those presented
for the trefoil knot. Figure \ref{fig:7} shows for example the
estimates of $t(q)$ for $4_{1}, 5_{1}$ and $5_{2}$  deep in the
adsorbed phase ($T=1.25$) compared with the one found for $3_{1}$.
All the curves look quite similar and by performing a linear
extrapolation  we obtain the estimates of $c$ given in table
\ref{table2}.

\begin{table}
\begin{center}
\begin{tabular}{|c|c|}
\hline
\bf{knot type} & \bf{c} \\
\hline
$3_1$    & $2.55 \pm 0.10$\\
$4_1$    & $2.60 \pm 0.10$\\
$5_1$    & $2.55 \pm 0.10$\\
$5_2$    & $2.56 \pm 0.16$\\
\hline
\end{tabular}
\caption{(color online) Estimates of the exponent $c$ for
different prime knots at $T=1.25$. } \label{table2}
\end{center}
\end{table}
The estimated exponents are well compatible with each other.
However, there seems to be a systematic difference between the
behavior of $3_{1}$ and $5_{1}$ and that of the other knots, as can be seen
from the plots in Figure \ref{fig:7}. 
Note that $3_{1}$ and $5_{1}$ belong to the
family of \emph{torus knots} \cite{Rolfsen} and it could be that, for knots
belonging to this family, $\langle\ell^{q}\rangle$ displays an $N$
behavior that is identical, not only in terms of the localization
exponent $t(q)$, but also in terms of the amplitude. The curve for
the knot $4_1$ is the most far apart and this is maybe due to the
fact that $4_1$ is achiral and belonging to a different family.

\subsection{Equilibrium behavior of the hosted knot.}
\begin{figure}[tbp]
\includegraphics[angle=0,width=\WIDTHE]{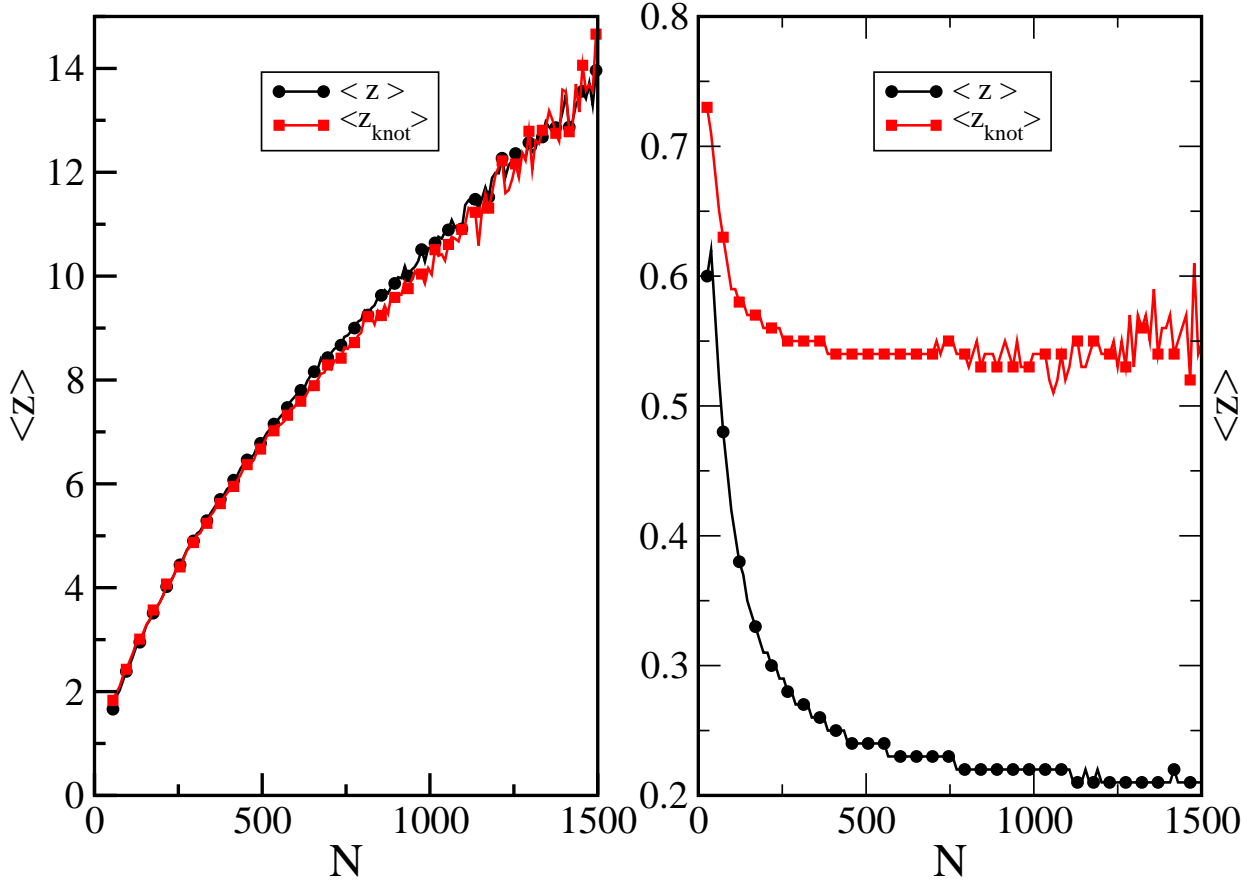}
\caption{(Color online) Average height $\langle z \rangle$ of the
SAP and of its knotted counterpart  ($\langle z_{knot} \rangle$)
as a function of $N$. The left
panel refers to $T=3.5$ (desorbed phase) while the right one to 
$T=1.25$ (strongly adsorbed phase). }
\label{fig:4}
\end{figure}

Having established that the adsorption transition drives knots
from being weakly localized to be strongly localized, it is now
interesting to understand whether the knot behaves as the rest of
the chain. In fact one may wonder if the typical equilibrium
configurations in the various phases are the ones in which the knotted part is expelled
out of the plane, so that the knot is free to fluctuate in the bulk. 
This can be checked by comparing for example  the average height
$\langle z \rangle$ of the whole SAPs, $\langle z \rangle$,
to the one restricted to its knotted part, $\langle z_{knot} \rangle$. 
Figure \ref{fig:4} shows the $N$ dependence of
$\langle z \rangle$, and $\langle z_{knot} \rangle$,
at two values of the temperature. At $T=3.5$ (left panel) the two
average heights are practically identical  suggesting that, above the
adsorption transition, the knotted part is indistinguishable from the
hosting ring. This behavior seems to change at $T=1.25$ (deep
adsorbed phase), where $\langle z \rangle$ is
systematically lower than  $\langle z_{knot} \rangle$.
This could indicate that in the adsorbed regime the knotted
part tends to be, on average, further away from the adsorbing plane
than  the whole chain. Note however that,
even in the strongly adsorbed phase, the knotted part
must keep a minimal number of excursions, in order to connect
the minimal number of crossings required by its topology. 
It turns out that this minimal number of excursions is sufficient to
explain the differences shown in Fig. \ref{fig:4} (right panel). 
This can be seen as follows.
By simulationing unknotted rings at $T=1.25$ with $N \simeq 76$ monomers 
(which is roughly the equilibrium length of the $3_{1}$ knot at that temperature), we observe that the average number of monomers in the
excursions is $\langle b\rangle \approx 32$. 
On the other hand, the knotted portion of a knotted ring at the same temperature
has $\langle b \rangle \approx 42$. 
In both cases, almost all excursions have height $z=1$. 
Hence the knot has, on average,  $\approx 10$ more  monomers in the bulk with
respect to tis unknotted counterpart. This is in agreement with the discrepancy 
observed in  Figure~\ref{fig:4} (right panel). This difference
is then due only to the unavoidable crossings pertaining to the knot. In 
fact each crossing requires at least 3 excursions of lenght $3$ each,
resulting to a minimal excursion length $3\cdot3=9\simeq10$ for the knotted
part.  Similar considerations apply to different knot types.
This suggests that the knotted parts of a polymer are essentially 
no more desorbed than the rest of the chain.

\begin{figure}[tbp]
\includegraphics[angle=0,width=\WIDTHE]{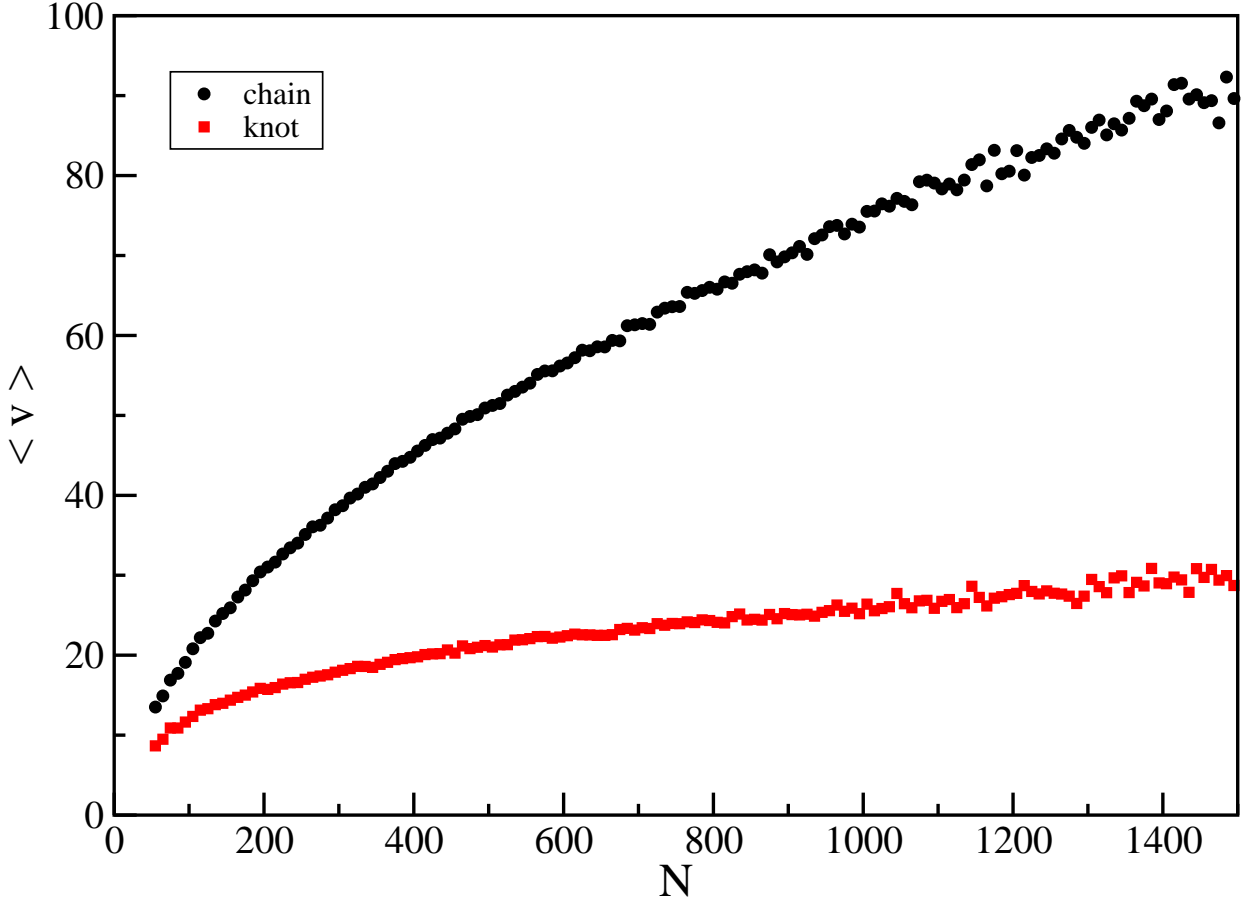}
\caption{Average number of visits $\langle v \rangle$  of the
whole SAP (top curve) and its knotted part (bottom curve) as a
function of $N$. The temperature considered is $T=3.50 \simeq
T_{c}$. The dashed line is still the one for the knotted one but
now multiplied by $(N/\ell)^{\phi}$.} \label{fig:6}
\end{figure}

Given that the position of the knot within the chain and with
respect to the adsorbing plane has nothing special if compared to
any remaining part of the chain, it is reasonable to expect that
the equilibrium properties themselves, once restricted to the
knotted part, will display the same features as those of the
whole chain. To confirm this picture we estimate the average
number of visits as a function of $N$ for the knotted
part and compare it with the one of the whole polygon (Figure \ref{fig:6}).
The temperature considered is the adsorption temperature $T_c$
where the scaling behavior (\ref{crossover}) with $\phi\approx
1/2$ is known to hold for \emph{all} rings (unrestricted
topology). By performing a simple power law fit we indeed get
$\phi = (0.53 \pm 0.02$) for the whole knotted SAP. Note that the estimate
is for SAPs with \emph{fixed knot type}, suggesting that the
crossover exponent could be unaffected by topological
constraints. To compute a crossover exponent restricted to the
knotted part, the scaling \ref{crossover} must be replaced by

\begin{equation}
\label{6} \langle v_{knot}\rangle \sim \langle \ell \rangle
^{\phi_{knot}},
\end{equation}

where $\langle v_{knot}\rangle$ indicates the number of visits of the knotted
part of the ring.
This gives  $\phi_{knot}=(0.49 \pm 0.04)$, compatible with
$\phi=\phi_{knot}=1/2$.

\begin{figure}[tbp]
\includegraphics[angle=0,width=\WIDTHE]{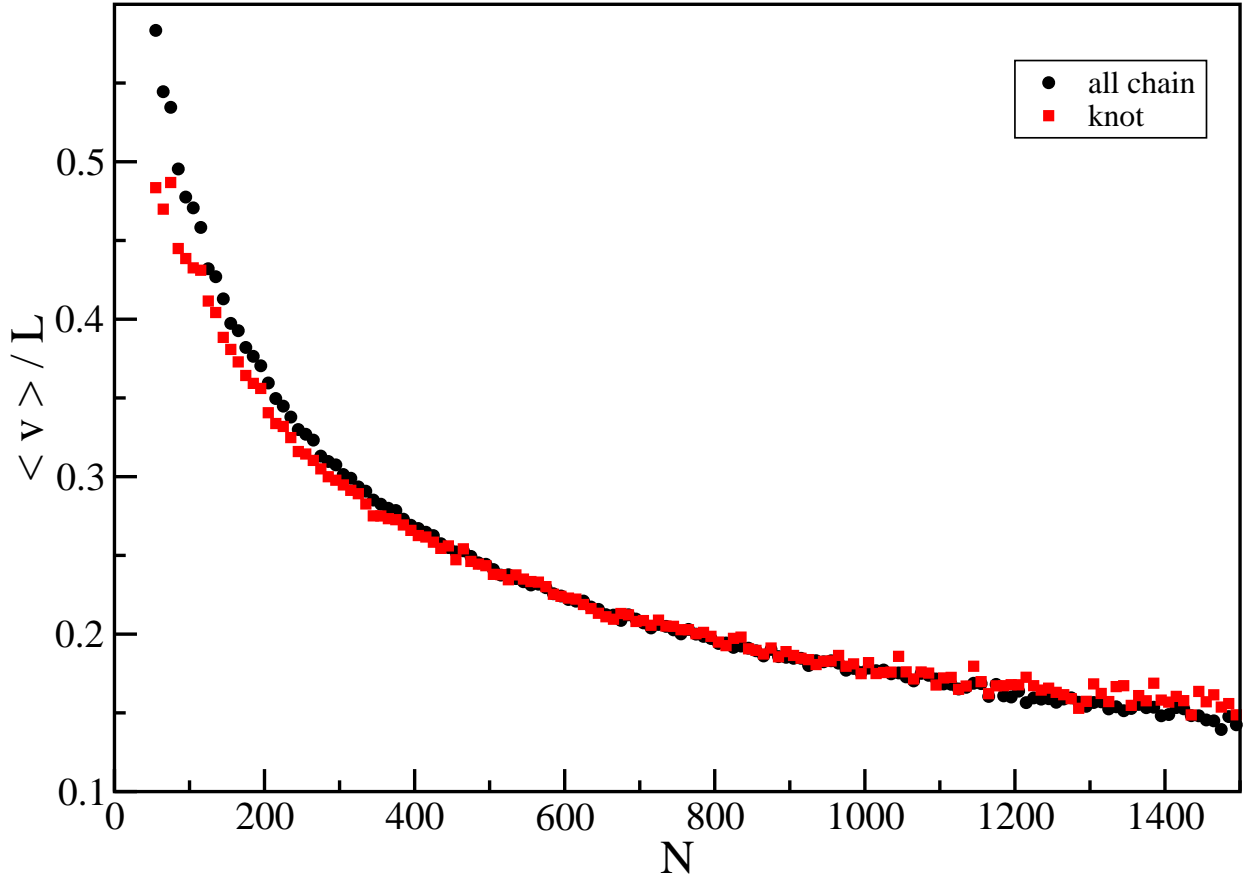}
\caption{Average number of visits of figure \ref{fig:6}
scaled by $L$ where $L=N$ for the whole SAP and $N^0.72$ for the knotted part.
}
\label{fig:6bis}
\end{figure}

Similarly one can define the metric exponent of the knotted
portion of the chain , $\nu_{knot}$ as

\begin{equation}
\langle R_{knot} \rangle \sim \langle \ell \rangle ^{\nu_{knot}}.
\label{rknot}
\end{equation}
The estimates are reported in Table \ref{table3} and compared with those
for the whole chain.

\begin{table}
\begin{center}
\begin{tabular}{|c|c|c|}
\hline
\bf{$T$} & \bf{$\nu$}& \bf{$\nu_{knot}$} \\
\hline
$\infty$ & $0.59 \pm 0.01$& $0.61 \pm 0.03$\\
$3.50$  & $0.60 \pm 0.01$& $0.64 \pm 0.03$\\
$2.75$  & $0.75 \pm 0.01$& $0.75 \pm 0.03$\\
$2.00$  & $0.75 \pm 0.02$& $0.77 \pm 0.04$\\
$1.50$  & $0.75 \pm 0.02$& $0.81 \pm 0.06$\\
$1.25$  & $0.74 \pm 0.02$& $0.81 \pm 0.06$\\
\hline
\end{tabular}
\caption{Estimates of the exponent $\nu$ for different values of
$T$ for trefoil knots. They have been obtained by a finite size
scaling analysis of equations (\ref{nu}) and (\ref{rknot}).}
\label{table3}
\end{center}
\end{table}

The $\nu_{knot}$ determinations are always slightly higher than those of $\nu$ and
have slightly larger error bars. This is due to two effects: $\ell$
varies on a smaller range than $N$, and the data for $R_{knot}$
are noisier too. The measured values of $\nu_{knot}$ are however
comparable, within error bars, with the values of $\nu$ for the
whole SAP. At \emph{very} low $T$, when knots become strongly
localized the error bars on the estimates of $\nu_{knot}$ are
quite big. This is mainly due to the relatively small values of
$\langle\ell\rangle$ (between 16 and 100) that do not allow a good
asymptotic analysis of (\ref{rknot}).

\subsection{Are adsorbed knots behaving as flat knots ?}
\begin{figure}[tbp]
\includegraphics[angle=0,width=0.98\WIDTHE]{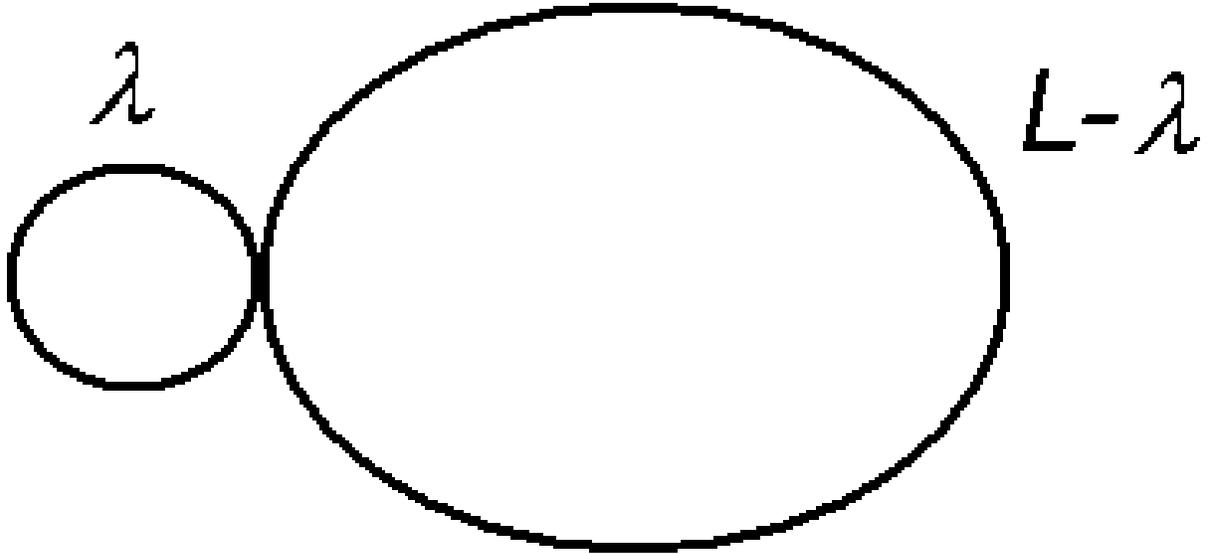}
\caption{Sketch of the figure-8 graph. Here $L$ is the total
length of the ring while $\lambda$ is the length of the
\emph{shorter} loop, that can be  identified with the knot length
$\ell$. It can be shown \cite{FlatKnots} that, when $L>>\lambda$, the pdf for
$\lambda$ scales as $\lambda^{-c_{8}}$, with $c_{8} = 2.69$.}
\label{fig:graf8}
\end{figure}

Our results show that the degree of localization of prime knots
appear to be independent on the knot type (see table
\ref{table2}). 
This is consistent with the theory of flat knots
which shows that, at leading order, all prime knots can be
asymptotically described by the \emph{figure 8} graph reported in
Figure \ref{fig:graf8},
whose behavior in the good solvent regime determines their strong localization
\cite{FlatKnots}. However,
adsorbed knots do not show the same degree of localization
(measured by the exponent $c$ presented above) as flat knots. In
fact, they are always somehow less localized, i.e. they show
values of $c$ lower than that provided by flat knot theory.  Note
however that for $T=1.25$ the $c$ values for adsorbed and flat
knots are comparable within 2 standard deviations. On the other
hand, since the Monte Carlo procedure deteriorates as $T$
decreases it would not be useful to try to go deeper in the
adsorbed phase. Indeed, there we where we would just estimate other values of $c$
that would possibly match better with the flat knot value, but with larger
error bars.

We can then  conclude that strongly adsorbed polymers should
eventually behave  as flat knots. If this happens  at some $T^{*}$
with $0<T^{*}<T_{c}$ or just in the limit
$T \rightarrow 0$ it is however impossible to establish
within our numerical precision.\\

Another distinction between adsorbed knotted rings and flat knots
concerns the number of crossings, $Cr$, the chain makes with itself
when it is \emph{regularly} projected on the adsorbing plane~\cite{nota6}.
Indeed, in the flat knot model the number of crossings is minimal ($3$ for the $3_1$,
$4$ for $4_1$, etc). This is crucial for the analytical treatment
of the model and it would be interesting to see how close is the behaviour of the
strongly adsorbed rings to this assumption.

It turns out that while the average number of crossings, $\langle
Cr \rangle$, of the \emph{whole chain} increases (approximately
linearly) with $N$, the same quantity restricted to the knotted
portion is approximatively constant and close to the minimum value
allowed by the topology. For example, $\langle Cr_{3_{1}}
\rangle\approx 3.6$. The independence of $N$ is quite
reasonable since deep into the adsorbed phase knots are strongly
localized. Moreover, strongly adsorbed
rings tend to minimize their energy by maximizing the number of
visits and, in the limit $T\to 0$, we should expect all the
unimportant crossings to disappear leaving a 2D ring with
just the essential (topological) crossings defining the knot \cite{nota3}. \\
At $T\ne0$, however, fluctuations in the number of crossings are
present and this could explain the deviation of the $c$ exponent
from the flat knot value for finite $T$ values. In this respect, it
could be useful to study the behavior of flat knots when the
restriction on the minimal crossing number is relaxed. Recently,
Guitter and Orlandini \cite{FlatKnotsGuitterOrlandini} introduced
a model of flat knots on a lattice where the ring is a 2D polygon with a
 number of crossings which can be changed and tuned with an
appropriate crossing fugacity. By implementing  our knot detection
technique on this model we have estimated the average knot size
for flat knots as a function of $N$ for a wide range of values of the average crossing number. 
We find that flat knots are always strongly localized, independently
of the average crossing number. This suggests that the crossing
number is not a key feature to explain the value of $c$ and that
the flat knot model, even with a fluctuating number of crossings,
is not a fair model for knotted rings in the adsorbed phase
unless $T\to 0$.\\

\section{Conclusions}

In this paper we study by Monte Carlo simulations the equilibrium
properties of self avoiding polygons with fixed topology (knot)
adsorbing onto an impenetrable wall. For unrestricted topology
it is known that SAPs undergo an adsorption transition from a
desorbed 3D swollen phase to an adsorbed phase. We first show that
for SAPs with fixed topology the adsorption transition is still
present. Moreover, we give numerical evidence that the metric
exponent $\nu$ in the adsorbed and desorbed phases and the crossover exponent
 $\phi$ for SAPs with
fixed topology, agree with the ones for the unrestricted topology
case. Even non-universal quantities such as the critical
adsorption temperature $T_c$ and the critical fugacity $K_c^o$ 
seem to be unaffected by the topological constraint.

By using a novel algorithm that allows the identification of the
knotted portion of the SAP \cite{OurShortKnotPaper} we are able to
focus on the equilibrium critical properties of this portion and compare them
with those the whole SAP. We show  that the knotted part
behaves, in most respects, as any other connected subset of the ring
with the same length. For example, we find that at $T=T_{c}$ the
average energy of the knot scales as $\langle \ell \rangle
^{\phi}$ where the value of $\phi$ agrees with that of the
whole ring. The metric exponent of the knot $\nu_{knot}$, which describes the
scaling of the radius of gyration of the knotted part of the ring as a function of
the knot length, is also consistent with the value of $\nu$ of the whole chain at every temperature.
Furthermore, the average displacement of the knot from the
plane (height) is the same of that of the whole ring, indicating
that there is not a preferred height in space for the knotted
portion. 

The main emphasis of our work  is however on the
 localization behavior of the knotted portion of the ring as
$T$ varies. We find that, for $T \ge T_{c}$, the average length
$\langle \ell \rangle$ of the knotted portion grows as $ N^{t}$
where $t\approx 0.72 $, consistent with the value found in
\cite{OurShortKnotPaper,OurLongKnotPaper} for knotted rings in the
3D bulk. This shows that knots in the desorbed phase and right at
the adsorption transition are weakly localized, i.e. the presence
of an attracting impenetrable plane does not change the
localization properties of the knot. Below $T_{c}$, the knot
becomes more and more localized reaching a strong localization
regime deep into the adsorbed phase. 
This crossover to more localized states is 
certainly triggered by the adsorption transition, but is quite smooth, as witnessed
by the $T$ dependence of the estimated localization exponent $t$. 
Thus, we can not exclude that below $T_{c}$ there is a 
\emph{continuous variation} of the exponent $t$ with $T$. 
A possible alternative scenario is the existence of a sharp
 localization transition at some 
$T_{loc}<T_{c}$, such that $t$ has the bulk value for $T>T_{loc}$, while $t=0$ 
(and $c=2.69$)) below. To justify this latter scenario, one must  assume that our data are
 affected by strong finite-size corrections, which cannot be numerically detected 
unless one performs simulations at much larger values of $N$. Note, however, that,
even under this assumption, the possibility that $T_{loc}=T_{c}$
should be discarded in account of the observation that $t$ starts to decrease
only below $T_{c}$; a crossover region is expected to be more symmetric 
around the transition temperature $T_{loc}$.
For sufficiently low
values of $T$ (deep in the adsorbed phase) the estimated $c$
exponent agrees, within error bars, with the one found for flat
knots. However, due to the large statistical uncertainty we cannot
rule out the possibility that the flat knot regime is reached only
in the limit $T \to 0$.

These results suggest that the relation between the flat knots
model and adsorbed knots is nontrivial. The flat knot regime is
reached only at low enough $T$ (or possibly at $T\to 0$ ) where the
number of crossings, arising from a regular projection on the
adsorbing plane, is close to the minimal value dictated by the topology. On the
other hand, in the whole adsorbed phase  many excursions in the
bulk are allowed and give rise to a number of crossings that exceed
the minimal one and that can fluctuate widely as $N$
increases. These fluctuations are certainly responsible for the deviations from flat-knot
behaviour observed at high enough $T$. The flat knot model is certainly not
an adequate representation of knotted rings in the whole adsorbed phase.

\section{Acknoledgment} This work was supported by FIRB01 and  MIUR-PRIN05.


\begin{thebibliography}{99}
\bibitem{note2} We recall that for adsorbing polymers $K_c = exp(\kappa)$ for all $T> T_c$.
\bibitem{notax} 
The standard error of our determinations of $K_c(\tau)$ and $K_c^o(\tau)$ is
of order $10^{-3}$.
\bibitem{note3} This assumption is made by analogy with the
known result for adsorbing SAPs with unrestricted topology
\cite{HammerTorrieWhitting}.
\bibitem{note4} Since for the distribution in Eq. \ref{scalP} it is found in
general that $t(q)=0$ for $q \leq q_c$ with $q_c>0$, the
normalization constraint $\int p(\ell,N)=1$ does not imply
$D(1-c)=0$ i. e. $c=1$.
\bibitem{nota6}
Note that a projection on the adsorbing plane ($z=0$) is not a
regular one. We then project the SAP on a plane that makes a small
angle with the $z=0$ plane.
\bibitem{nota3}The reader should keep in mind that the polymer
makes excursions outside the $xy$ plane for every $T>0$.
\end{thebibliography}
\end{document}